\documentclass[12pt,a4paper]{article}

\usepackage[scale={.75,.8}]{geometry} 
\usepackage{pstricks}
\usepackage[dvips]{graphicx}
\usepackage[latin2]{inputenc}
\usepackage{epic}
\usepackage{latexsym}
\usepackage{epsf}
\usepackage{epsfig}
\usepackage{amssymb,amsmath}

\newcommand{\eref}[1]{eq.\ (\ref{e.#1})}

\newcommand{\cref}[1]{Chapter \ref{c.#1}}

\def\nn{\nonumber \\}

\def\beq{\begin{equation}} 
\def\eeq{\end{equation}} 
\def\bea{\begin{eqnarray}}  
\def\eea{\end{eqnarray}}  
\newcommand{\bal}{\begin{align}}
\newcommand{\eal}{\end{align}}   
\def\ba{\begin{array}}  
\def\ea{\end{array}}   
\def\bi{\begin{itemize}}  
\def\ei{\end{itemize}}  
\def\ben{\begin{enumerate}}  
\def\een{\end{enumerate}}  
\def\beq{\begin{equation}}  
\def\eeq{\end{equation}}  
\def\bc{\begin{center}}
\def\ec{\end{center}} 
 \def\bt{\begin{table}}
\def\et{\end{table}}  
 \def\btb{\begin{tabular}}
\def\etb{\end{tabular}}  
\newcommand{\bvec}{\left ( \ba{c}}
\newcommand{\evec}{\ea \right )}

\def\cl{{\mathcal L}}

\def\co{{\mathcal O}}

\def\gev{\, {\rm GeV}}

\def\mass2{mass${}^2$}

\newcommand{\ms}{M_{\rm soft}}

\def\ra{\rangle}
\def\la{\langle}  

\def\simgt{\stackrel{>}{{}_\sim}}

\newcommand{\ti}{\tilde}

\def\ov{\overline}

\begin{document}

\pagestyle{empty}
\begin{flushright}
CERN-PH-TH/2006-051\\ 
{\bf \today}
\end{flushright}
\vspace*{5mm}
\begin{center}

{\large {\bf Twin SUSY }}\\
\vspace*{1cm}

{\bf Adam~Falkowski}$^{\rm a),b)}$\footnote{Email:adam.falkowski@cern.ch},
{\bf Stefan~Pokorski}$^{\rm b)}$\footnote{Email:pokorski@fuw.edu.pl}
and 
{\bf Martin~Schmaltz}$^{\rm c)}$\footnote{Email:schmaltz@bu.edu}
\vspace{0.5cm}
 
a) CERN Theory Division, CH-1211 Geneva 23, Switzerland\\
b) Institute of Theoretical Physics, Warsaw University, Ho\.za 69, 00-681 Warsaw, Poland\\
c) Physics Department, Boston University, Boston, MA 02215, USA.

\vspace*{1.7cm}
{\bf Abstract}
\end{center}
\vspace*{5mm}
\noindent
{ 
We construct an extension of the MSSM in which superpartners can naturally be
heavier than the electroweak scale. This ``little hierarchy'' of scales is 
stable because the Higgs arises as a pseudo-Nambu-Goldstone boson in the
breaking of an accidental $SU(4)$ symmetry of the Higgs sector.
Supersymmetry and the global symmetry combine to forbid logarithmically
divergent one-loop contributions to the Higgs mass.
The accidental symmetry follows from a simple ``twin'' parity which exchanges the
$SU(2)$ sectors in the $SU(3)_C\times SU(2)_L \times SU(2)_R \times U(1)_X$
gauge group.
}
\vspace*{1.0cm}
\date{\today}


\vspace*{0.2cm}
 
\vfill\eject
\newpage

\setcounter{page}{1}
\pagestyle{plain}

\section{Introduction}

Supersymmetry is a very attractive scenario for physics at the TeV scale. Unfortunately,
it's simplest implementation, the MSSM, requires fine-tuning of parameters
once experimental constraints are imposed. The problem can be summarized as follows:
In the MSSM scalar superpartners and the Higgs scalars are on equal footing,
leading us to expect that the superpartner mass scale $M_{SUSY}$ is equal
to the Higgs mass and the electroweak scale $M_{W}\sim 100$ GeV.
Clearly, this expectation is not borne out in Nature as shown
by direct searches, limits from precision electroweak and flavor constraints, and
from the lower bound on the mass of the Higgs boson. Instead the data prefer
that the superpartners are heavier than the electroweak scale. The next-most natural
expectation might be $M_{SUSY} \sim 4\pi\, M_{W}$, and we must ask if this modest hierarchy of
scales is stable under radiative corrections.

In the MSSM, 
the largest radiative correction to the Higgs soft mass is due to top/stop loops
\beq
\label{e.mssmhmp}
\delta m_H^2 \simeq
 - \frac{3 y_t^2}{8 \pi^2}\,  m_{\tilde t}^2\, \log \left( \frac{\Lambda_{UV}^2}{m_{\tilde t}^2} \right)
\eeq
where $y_t$ is the top Yukawa coupling,  $m_{\tilde t}$ is the stop mass (for simplicity, we consider
degenerate left and right-handed stops and ignore stop mixing),
and $\Lambda_{UV}$ is the high energy scale at which the
soft masses are generated. We see that for large values of $\Lambda_{UV}$,
for example $\Lambda_{UV}\sim M_{Planck}$ in gravity mediation,
the suppression due to the loop factor is canceled by
the large logarithm, and a hierarchy between superpartner masses and the Higgs mass
is unstable.

Improving naturalness of supersymmetric theories with heavy superpartners
therefore requires removing the large logarithm.%
\footnote{Increasing the Higgs quartic coupling with new contributions
from an extended Higgs sector without reducing the soft mass leads
to new fine tuning problems as was recently emphasized in
\cite{schuto}.}
Two possibilities suggest themselves.
One  is to lower the scale $\Lambda_{UV}$ down to 1-100 TeV range. 
This can be achieved in scenarios in which supersymmetry breaking is mediated at a low scale,
such as gauge mediation or theories with large extra dimensions.  
Effectively small $\Lambda_{UV}$ may also be a consequence of an interplay between
gravity and anomaly mediation, as recently pointed out in \cite{chjeko}. 
But even in a theory in which such a low mediation scale is realized one still wonders
what makes the Higgs different, i.e. why is the Higgs soft mass much lower than
typical superpartner masses?
We do not pursue this avenue in this paper and instead allow that TeV size superpartner masses
are generated at high scales. 

The other possibility is to treat the Higgs differently from the superpartners
by making it a pseudo-Nambu-Goldstone boson as in little Higgs theories \cite{lh} but
now in the context of supersymmetry \cite{chacko,chfapo} and \cite{bechfa,rosc,csmash}.
In this approach, radiative corrections to the Higgs soft mass are finite because they
are ``doubly protected'' by softly broken supersymmetry and by the little Higgs mechanism. 
At tree-level the soft mass of a doubly protected Higgs vanishes, 
while the dominant radiative correction has the form 
\beq
\delta m_H^2  \approx  - {3 y_t^2\over 8 \pi^2} \left [ 
 ( m_{\tilde t}^2 + m_T^2) \log ( m_{\tilde t}^2 + m_T^2) -  
 m_{\tilde t}^2 \log (m_{\tilde t}^2) - m_T^2 \log (m_T^2) \right ] \, ,
\eeq
where $m_T$ is the mass of the little Higgs partner of the top.
In the limit in which the little Higgs partners are much heavier than the superpartners this
formula reduces to \eref{mssmhmp} with $\Lambda_{UV}$ replaced by $m_T$.

While the idea of double protection is very simple,
explicit implementation encounters several problems. 
The original model of Birkedal {\it et. al.} \cite{chacko} relied on
enlarging the Standard Model $SU(2)_{weak}$ into a {\it global} $SU(3)$
symmetry by introducing Higgs and top partners which complete $SU(3)$ triplets.
The problem with this approach is that the Standard Model gauge interactions
do not respect the global $SU(3)$ symmetry.
Renormalization group running  from $\Lambda_{UV}$ down to the weak
scale badly breaks the $SU(3)$ symmetry so that it can no longer protect the soft Higgs mass. 
Note that this problem affects models which rely on a global symmetry
which is explicitly broken by gauge interactions (e.g. a SUSY version of the littlest Higgs).

More recent attempts at implementing double protection therefore extend the
global symmetry to the gauge interactions.
In refs. \cite{bechfa,rosc,csmash} the gauged $SU(2)_{weak}$ is enlarged to a
gauged $SU(3)_{weak}$ as in the simplest little Higgs \cite{simplest}.
The pseudo-Nambu-Goldstone bosons (pNGBs) then arise because the $SU(3)_{weak}$ gauge
symmetry is broken spontaneously to $SU(2)$ by two different sets of fields. 
If the coupling between these two sets of fields is sufficiently weak,
then the theory has an approximate $SU(3)^2$ symmetry which is spontaneously broken to $SU(2)^2$,
yielding two sets of pNGBs, one linear combination is eaten by the heavy $SU(3)_{weak}$
gauge bosons, the other remains light.
A general problem with this approach is that the $SU(3)_{weak}$ D-terms
strongly couple the two sectors and explicitly break the two $SU(3)$'s of the
Higgs sector to a single $SU(3)$.
The would-be pNGBs get a mass from this D-term, and more model-building is required to prevent it. 
Therefore realistic models end up being rather complicated.
 
In this paper we try a new approach based on the twin Higgs idea \cite{chgoha1,chgoha2}
in which the global symmetry protecting the Higgs potential arises as an accidental
symmetry after imposing a much more modest $Z_2$ twin parity.
This spontaneously broken accidental symmetry makes the Higgs a pNGB and
ensures double protection. 
An advantage of this idea is that it is relatively  easy to implement the $Z_2$
symmetry by enlarging the field content of the MSSM.

The model we study first is left-right symmetric with the gauge group
$SU(3)_C \times SU(2)_L \times SU(2)_R \times U(1)_X$ as in \cite{chgoha2}.
The $Z_2$ symmetry interchanges the left and right $SU(2)$ gauge bosons.
Furthermore, every MSSM field has its $Z_2$ partner. 
In particular, the Higgs sector consist of  four multiplets: 
two ``left'' doublets $H_u$ and $H_d$ and two ``right'' doublets  $\ti H_u$ and $\ti H_d$. 
The $Z_2$ symmetry imposed on the Higgs sector is sufficient to guarantee an
accidental $SU(4)$ symmetry of the dimension 2 terms in the Higgs potential.
Thus even though the Yukawa interactions which renormalize the Higgs mass terms
do not respect the full $SU(4)$ (even after imposing the $Z_2$ symmetry), the resulting
corrections to the Higgs masses are automatically $SU(4)$ symmetric. 
This is how double protection is realized in this model. Divergent radiative
corrections to soft masses do not lead to masses for the pNGB because they respect
the full global symmetry. 

Unfortunately, this minimal twin supersymmetric model shares a problem
with the models based on $SU(3)_{weak}$ group discussed above: some of the quartic
couplings in the Higgs sector, in particular the  $SU(2)_L \times SU(2)_R \times U(1)_X$ D-terms
explicitly break the $SU(4)$ symmetry and lead to large tree-level masses
for the would-be pNGBs. We demonstrate this problem in the minimal twin susy model in
the next Section. In Section 3 we show that a simple modification of the model
can avoid the troublesome D-term contributions and lead to a fully realistic model. 
Section 4 contains our conclusions.

\section{A Toy Model for Twin SUSY}

Our toy model, inspired by ref. \cite{chgoha2}, is left-right symmetric with the gauge group  
$SU(3)_C \times SU(2)_L \times SU(2)_R \times U(1)_X$.
An additional discrete $Z_2$ symmetry interchanges the left and right $SU(2)$ gauge groups,
thus enforcing $g_L  = g_R$.  
The discrete symmetry requires that for every ``left'' doublet there
exists a $Z_2$ partner (denoted by tilde) that transforms as a doublet under $SU(2)_R$. 
Therefore the minimal Higgs sector contains four multiplets: 
two left doublets $H_u$ and $H_d$ (required by anomaly cancellation) and
two right doublets  $\ti H_u$ and $\ti H_d$ (required by the $Z_2$).
Their $U(1)_X$ charges are chosen as $+\frac12$, $-\frac12$, $-\frac12$ and $+\frac12$, respectively.   
With a help of a singlet superfield $N$ we can write a superpotential that yields
interactions between the left and right Higgs doublets, 
\beq
\label{e.snh}
W  = \lambda N( H_u H_d + \ti H_u \ti H_d - F^2)
\eeq 
Note that the $Z_2$ symmetry ensures an accidental global
$SU(4)$ symmetry\footnote{%
Emergence of SU(4) symmetry in such setup was also noted in ref. \cite{zurab1}.}
 under which $(H_u,\ti H_u)$ transform as
$\bf \bar 4$ and $(H_d,\ti H_d)$ as $\bf 4$. 
Furthermore, it requires the soft mass terms of the Higgses to be $SU(4)$ symmetric
\beq
\cl_{soft} = - M_u^2(|H_u|^2 + |\ti H_u|^2)  - M_d^2(|H_d|^2 + |\ti H_d|^2) \ .
\eeq

The linear term in \eref{snh} forces one of the Higgs pairs (which we assume to
be $\ti H_u$, $\ti H_d$) to acquire a vev,
$\la \ti H_u \ra  = f \sin \beta $, $\la \ti H_d \ra = f \cos \beta$, 
where $f$ depends on $F$ and the soft masses and
$\tan^2\!\beta = (M_d^2 + \lambda^2 N^2)/(M_u^2 + \lambda^2 N^2)$.  
The $SU(4)$ symmetry is spontaneously broken down to $SU(3)$ yielding $7$
Nambu-Goldstone bosons. Three are eaten as a result of gauge symmetry breaking 
$SU(2)_R\times U(1)_X \to U(1)_Y$, leaving four physical NGBs. 
These form an $SU(2)_L$ doublet $H$ that is identified with the SM Higgs field.  
In the non-linear sigma model parametrization:
\bea
\label{e.gp}
 H_u \to  f  \sin\! \beta\,  \sin ({|H|/f})\ {H \over |H|} &\qquad & 
  H_d \to  f \cos\! \beta\,  \sin ({|H|/ f})\ {\epsilon H^* \over |H|}
 \nn
 \ti H_u \to  f \sin\! \beta\,  \cos ({|H|/ f})\ \bvec\! 0 \!\\\! 1 \!\evec    &\qquad &  
\ti H_d \to   f \cos\! \beta\,  \cos ({|H|/f})\   \bvec\! 1 \!\\ \!0 \!\evec 
\eea 
Note that $\tan\!\beta$ for the MSSM Higgs fields $H_u$ and $H_d$ is equal to 
the ratio of the VEVs of the heavy Higgses $\ti H_d$ and $\ti H_u$. 
With this parametrization it is easy to verify explicitly that neither the F-term
potential nor the soft terms depend on the SM Higgs field $H$. 

On the other hand, gauge and Yukawa interactions break $SU(4)$ explicitly,
even after imposing the $Z_2$ symmetry.
Therefore these interactions  generate a potential for the SM Higgs at loop level.   
However,  the $Z_2$ is sufficient to ensure double protection.
The point is that $Z_2$ implies $SU(4)$ symmetry of all quadratic terms. 
Since the coefficient of the UV logarithm in the loop induced Higgs potential
is quadratic in the Higgs fields (and quadratic in the supersymmetry breaking
mass parameters), we are guaranteed it does not depend on the SM Higgs field $H$.

Renormalizable and $Z_2$ symmetric
Yukawa interactions can be realized in our setup by
introducing $Z_2$ partners for each MSSM matter multiplet
(including ``right-handed'' neutrinos). 
The $SU(3)_C \times SU(2)_L \times SU(2)_R \times U(1)_X$ representation of the MSSM
fields and their partners are given by (see also \cite{zurab2}):
\bea
\label{e.particles}
Q \to (3,2,1)_{1/6} & \qquad & \ti Q^c \to (\bar 3,1,2)_{-1/6}
\nn
L \to (1,2,1)_{-1/2} & \qquad &  \ti L^c \to (1,1,2)_{1/2}
\nn
T^c \to (\ov 3,1,1)_{-2/3} & \qquad &  \ti T \to (3,1,1)_{2/3}
\nn
B^c \to (\ov 3,1,1)_{1/3} & \qquad &  \ti B \to (3,1,1)_{-1/3}
\nn
\tau^c \to (1,1,1)_{1} & \qquad &  \ti \tau \to (1,1,1)_{-1}
\nn
N_\tau^c \to (1,1,1)_{0} & \qquad &  \ti N_\tau \to (1,1,1)_{0}
\eea 
where the $U(1)_X$ charges are determined from the corresponding hypercharges
using the identification $Y=X-T^3_R$. The Yukawa interactions can be written as:
\bea
\label{e.yuks}
W& =  & Y^t H_u Q T^c + Y^t \ti H_u \ti Q^c \ti T  + M^t \ti T T^c 
\nn
&+  & Y^b H_d Q B^c + Y^b \ti H_d \ti Q^c \ti B  + M^b \ti B B^c 
\nn
&+  & Y^\tau H_d L \tau^c + Y^\tau \ti H_d \ti L^c \ti \tau  + M^\tau \ti \tau \tau^c 
\nn
&+  & Y^n H_u L N_\tau^c + Y^n \ti H_u \ti L^c \ti N_\tau  + M^n \ti N_\tau N_\tau^c 
\eea 

We now compute the one loop contribution to the SM Higgs potential due to the top sector. 
For simplicity, we assume degenerate soft masses for stops ($m_{\ti t}^2$)
and vanishing A-terms
\bea
\label{e.dphp}
\delta V &=& \delta m_H^2 |H|^2 + \delta \lambda |H|^4 + \dots
\nn
\delta m_H^2  & \approx & - {3 \over 8 \pi^2} y_t^2 \left [ 
 (m_{\ti t}^2 + m_T^2) \log (m_{\ti t}^2 + m_T^2)
- m_{\ti t}^2 \log (m_{\ti t}^2) - m_T^2 \log (m_T^2) \right ] 
\nn
\delta \lambda &\approx& {3 \over 16 \pi^2}  y_t^4 \left [
\log \left ({ m_T^2 m_{\ti t}^2 \over (m_{\ti t}^2 + m_T^2) m_t^2 } \right )
 + {3 \over 2} - 2 {m_{\ti t}^2 \over m_T^2}
\log \left ({m_{\ti t}^2 + m_T^2 \over m_{\ti t}^2}  \right ) 
 \right ]
 \eea 
Here $m_T^2 = (Y^t f \sin\! \beta)^2 + (M^t)^2$ is the mass squared of the heavy top partner
and $y_t  = (Y^t \sin \beta)^2 f/m_T$ is the SM top Yukawa coupling.
These formulas are valid up to $\co(v/\!f)^2$ corrections.   
Note that the contribution to the SM Higgs mass is negative and
can trigger electroweak symmetry breaking.
The correction to the quartic terms is similar as in the MSSM, for
large $m_T^2$ or large $m_{\ti t}^2$ it becomes
$\delta \lambda \rightarrow (3 y_t^4/ 16 \pi^2)\,\log [{\rm Min}( m_{\ti t}^2,m_T^2)/ m_t^2]$.

Unfortunately, this nice and economical model is not viable.
The problem is very similar to that which arises in
gauged $SU(3)_{weak}$ models \cite{bechfa,rosc,csmash}.   
The $SU(2)_L \times SU(2)_R \times U(1)_X$ D-term potentials do not respect
the $SU(4)$ global symmetry and, in general, yield a tree-level  mass for the SM Higgs.
The effects of $SU(4)$ breaking are easy to understand in the limit $F > \ms$
which is required by lower bounds on the masses of the $SU(2)_R \times U(1)_X$ gauge bosons from LEP
experiments.
After integrating out the fields with masses of order $F$
and to lowest order in $\ms^2/F^2$
the potential for the SM Higgs has the form $V = V_{soft} + V_{quartic}$ where
\bea
\label{e.etp}
V_{soft} & = & |H|^2 \left[  
{M_u^2+ M_d^2 \over 2} (\cos\!\beta - \sin\! \beta)^2 
 + {M_u^2 - M_d^2 \over 2}(1+{g_Y^2 \over g_L^2})\,\cos^2 2\beta \right ]
\nn
V_{quartic} & = & |H|^4 \left[ {g_L^2 + g_Y^2 \over 8}\, \cos^2 2\beta \right]
\eea
We see that avoiding a large tree level soft mass for the Higgs requires $M_u^2 = M_d^2$ and,
in consequence, $\sin\!\beta = \cos\!\beta$, i.e. $\tan\!\beta =1$. 
But then the tree level quartic vanishes as well which is in
conflict with the lower bound on the Higgs mass from LEP2. The vanishing of the Higgs soft
mass at $\tan\!\beta =1 $ can be understood by noting that for $\tan\!\beta =1$ the
D-terms have vanishing expectation values. Therefore the D-term potentials are supersymmetric
and can only give a mass to the superpartners of the eaten NGBs, but not to the light Higgs doublet.

\section{A Realistic Model}

The problem with the toy model of the previous section is that the $SU(2)_R \times U(1)_X$
D-terms do not respect the global $SU(4)$ which protects the Higgs from obtaining
a mass. This implies that we either have a large soft mass for the Higgs ($\tan\!\beta\ne 1$)
or a vanishing tree level quartic for the Higgs ($\tan\!\beta= 1$).
Either choice is problematic. 

To solve this problem, two basic strategies are possible. 
One is to choose $\tan\!\beta\ne 1$ and remove the offending D-terms by breaking
the $SU(2)_R \times U(1)_X$ gauge symmetry at a higher scale
with additional charged fields.\footnote{%
Alternatively, the D-terms can be removed with ``supersoft'' \cite{fonewe} Dirac mass terms
for the $SU(2)_R \times U(1)_X$ gauginos \cite{chwe}.}. 
Alternatively, we may enforce $\tan\!\beta= 1$ with an approximate symmetry,
and obtain a new tree level contribution to the
Higgs quartic from appropriate superpotential couplings. 
This is the route which we take in the following.

We wish to stabilize the vevs of $\ti H_{u,d}$ at  $\tan\! \beta\! =\! 1$ which
requires $M_u^2\approx M_d^2$ because
$\tan^2\!\beta = (M_d^2 + \lambda^2 N^2)/(M_u^2 + \lambda^2 N^2)$.
We therefore impose an approximate $Z_2$ symmetry acting as $u\leftrightarrow d$.
Contributions to the Higgs mass from \eref{etp} only appear at second order in
the difference $M_u^2- M_d^2$ so that small radiative corrections to $M_u^2= M_d^2$
can be tolerated. In general, these corrections can be large due to one loop 
diagrams involving the Yukawa couplings in the top and bottom sector.
However, the $u\leftrightarrow d$ symmetry also implies $Y^t \approx Y^b$
in which case the radiative corrections respect $M_u^2= M_d^2$. Note that
the top and bottom quark masses can be split by choosing $M^b \gg M^t$ in \eref{yuks},
this breaks $u\leftrightarrow d$ softly, and a small Higgs mass is generated.

Since $\tan \beta = 1$ we need an additional contribution to the SM Higgs quartic term. 
Consider the model of the previous section with four additional singlets $S, \ti S, S^c, \ti S^c$
and the superpotential
\bea
W &=&  \kappa S (H_u H_d - M_s S^c) +  \kappa \ti S (\ti H_u \ti H_d - M_s \ti S^c) 
\label{e.extrasuper}
\eea
The purpose of the first superpotential term is to add a contribution to the
quartic coupling of the Higgs from the F-term of $S$.
In absence of a soft mass, the field $S^c$ would adjust its VEV
such as to cancel the quartic. We therefore require a sizable soft mass for $S^c$.
The second superpotential term is required by the $Z_2$ symmetry. When expanded in
terms of the light Higgs field this term can contribute a large soft mass to the Higgs.
We therefore require a small soft mass for $\ti S^c$ so that its VEV can adjust to
cancel the Higgs soft mass. Thus for our model to work we must explicitly break the $Z_2$
symmetry with the soft terms for $S^c$ and $\ti S^c$ (alternatively, the breaking
could be achieved by splitting the masses $M_s$ and $M_{\ti s}$). This explicit breaking
of the $Z_2$ in the soft masses is radiatively stable because the soft terms
for $S^c$ and $\ti S^c$ do not run at any loop order.

To determine the range of viable soft masses for $S^c$ and $\ti S^c$ we write the
contribution to the Higgs potential from \eref{extrasuper} which involves $S^c$ and $\ti S^c$
and their soft masses
\bea
V&=&\kappa^2 \left|M_s  S^c - {f^2\over 2} \sin({|H| \over f})\right|^2 +  m^2 |S^c|^2  \nn
 &+&\kappa^2 \left|M_s \ti S^c - {f^2\over 2} \cos({|H| \over f})\right|^2 + \ti m^2 |\ti S^c|^2
\eea
and minimize the potential for the singlets. We find that a sufficiently large quartic
requires $\kappa =\co( 1 )$, $m\simgt M_s$ and $\ti m \ll M_s$.
In this limit, the expressions for the Higgs soft mass and quartic are
\bea
m_H^2 &=& - \ti m^2\, {f^2 \over 2 M_s^2} \nn
\lambda &=& \frac{\kappa^2}4\,
{m^2 \over \kappa^2 M_s^2 + m^2} 
\eea
Taking for example $\kappa^2 = 1/2$, $M_s = m = 1$ TeV, $f=3$ TeV and $\ti m = 50$ GeV gives
a sufficiently large tree level quartic ($\lambda=1/12$) and a contribution to the soft mass
of order the Z-mass ($m_H = 106$ GeV). 
Note that at the minimum of the potential $S$ and $N$ do not have scalar
expectation values but that the $F$ component of $S$ is non-zero. 
Therefore \eref{extrasuper} contains an effective $B\mu$ term for the Higgs
but no $\mu$ term which is necessary in order to make the Higgsinos in $H_u$ and $H_d$ sufficiently heavy. 
To generate a Higgsino mass we add the $Z_2$ (and $SU(4)$) symmetric term
$\mu (H_u H_d + \ti H_u \ti H_d)$.  
Since the singlet $N$ has a soft mass term its vev cannot adjust to cancel the
$\mu$-term completely and Higgsino masses are generated.

\section{Unification}

The particle spectrum in \eref{particles} is suggestive of various steps of unification.
First quarks and leptons may be unified into an $SU(4)$ Pati-Salam ``color''
group where the $U(1)_X$ charges arise as a combination of the ${diag}(1,1,1,-3)$ generator
of $SU(4)$ and ``middle'' $U(1)_M$. 

The unified fields transform under
$SU(4)\times SU(2)_L \times SU(2)_R \times U(1)_M$ as
\bea
\label{e.particles1}
(Q,L) \to (4,2,1)_{0} & \qquad & (\ti Q^c,\ti L^c) \to (\bar 4,1,2)_{0}
\nn
(T^c,N_\tau^c) \to (\ov 4,1,1)_{-1/2} & \qquad &  (\ti T,\ti N_\tau) \to (4,1,1)_{1/2}
\nn
(B^c,\tau^c) \to (\ov 4,1,1)_{1/2} & \qquad &  (\ti B,\ti \tau) \to (4,1,1)_{-1/2}
\nn
H_u \to (1,2,1)_{1/2} & \qquad &  \ti H_u \to (1,1,2)_{-1/2}
\nn
H_d \to (1,2,1)_{-1/2} & \qquad &  \ti H_d \to (1,1,2)_{1/2}
\eea 
This particle content suggests a further unification where $U(1)_M$ is the
$T^3$ generator of a new $SU(2)_M$ with the field content
$SU(4)\times SU(2)_L \times SU(2)_M \times SU(2)_R$
\bea
\label{e.particles2}
\Psi_L=(Q,L) \to (4,2,1,1) & \qquad & \Psi_R^c=(\ti Q^c,\ti L^c) \to (\bar 4,1,1,2)
\nn
\Psi_M^c=(T^c,N_\tau^c,B^c,\tau^c) \to (\ov 4,1,2,1) & \qquad & 
\Psi_M=(\ti T,\ti N_\tau,\ti B,\ti \tau) \to (4,1,2,1)
\nn
H=(H_u,H_d) \to (1,2,2,1) & \qquad &  \ti H=(\ti H_u,\ti H_d) \to (1,1,2,2)
\eea 
The Yukawa couplings now take a particularly simple form
\bea
\label{e.gutyuk}
Y^L\,\, \Psi_L H \Psi_M^c + M\, \Psi_M \Psi_M^c + Y^R\,\, \Psi_M \ti H \Psi_R^c \ .
\eea

The ``unification'' presented here may also help in solving a problem with Landau
poles for the $SU(3)_c\times U(1)_X$ gauge couplings. With all ``twin'' partners
added, neither $SU(3)_c$ nor $U(1)_X$ are asymptotically free. Their respective
Landau poles occur at  $10^{13} \gev$ and $10^{11} \gev$
(if all extra fermion masses are at the TeV scale).
Above $SU(4)\times SU(2)_M $ breaking scale the running can be much slower
due to the contribution of extra gauge bosons,
provided the number of extra matter multiplets is small enough. 
Thus the Landau pole can be avoided assuming an $SU(4)\times  SU(2)_M $ breaking
scale well below $10^{11} \gev$. Note that this does not imply problems with proton decay
as ``unification'' into $SU(4)$ does not violate baryon number.
The situation here is somewhat better than in the $SU(3)_{weak}$ models,
where perturbative unification is difficult to achieve \cite{csmash}. 
However, the gauge couplings do not unify with the particle
spectrum in \eref{particles2}, it would require additional  matter multiplets.  
Furthermore, splitting of quark and lepton masses within generations is non-trivial,
especially for $\tan \beta \approx 1$.

\section{Conclusions}

In this paper we constructed a supersymmetric model in which the Higgs is a
pseudo-Nambu-Goldstone boson.
Our model is left-right symmetric with the electroweak gauge group extended to 
$SU(2)_L \times SU(2)_R \times U(1)_X$.
Imposing a $Z_2$ twin parity which interchanges the ``left'' and ``right'' $SU(2)$'s implies 
an accidental $SU(4)$ symmetry of all mass terms in the extended Higgs sector.
The Higgs pNGB arises after spontaneous breaking of this $SU(4)$.
The interplay of the global symmetry and supersymmetry leads to double protection: 
the Higgs  mass parameter does not receive logarithmically divergent corrections
at one loop, even in the presence of soft supersymmetry breaking.
The reason is that at one loop only dimension two terms can be logarithmically
renormalized, but these are automatically $SU(4)$ symmetric by virtue of the $Z_2$.   

The purpose of this construction is a solution to the supersymmetric little hierarchy problem.
Removing the one-loop logarithmic divergence allows for $M_{SUSY} \sim 4\pi\, M_{W}$ without fine-tuning.
Our model provides an explicit realization of double protection
which is stable under radiative corrections. 
Technical complications related to  the $SU(2)_R \times U(1)_X$ D-terms which
do not respect the global symmetry and contribute to the pNGB mass at tree-level
force us to introduce an extended Higgs sector. Thus our solution to the little
hierarchy problem in the MSSM comes at the price of simplicity - several new gauge
singlets with carefully designed interactions are required at the TeV scale. 
The idea of stabilizing little hierarchy by global symmetries deserves further study,
both at the theoretical and at the phenomenological level.

\section*{Acknowledgments}

M.S. thanks CERN Theory for hospitality and T. Roy for comments on
the manuscript.
A.F. and S.P. were partially supported by the European Community 
Contract MRTN-CT-2004-503369 for years 2004-2008 and by the Polish KGB
grant 1 P03B 099 29 for years 2005--2007. 
M.S.  acknowledges support from an Alfred P. Sloan Research Fellowship
and DOE grant DE-FG02-91ER40676.


\end{document}